\documentclass[aps,prd,eqsecnum,11pt,showpacs,preprintnumbers,nofootinbib]{revtex4-1}
\usepackage{amsmath,amsfonts,amssymb}
\usepackage{graphicx, tikz}
\usepackage{bm,color}
\newcommand{\be}{\begin{equation}}
\newcommand{\ee}{\end{equation}}
\newcommand{\bes}{\begin{subequations}}
\newcommand{\ees}{\end{subequations}}
\newcommand{\bea}{\begin{eqnarray}}
\newcommand{\eea}{\end{eqnarray}}
\newcommand{\la}{\langle}
\newcommand{\ra}{\rangle}
\def\frc#1#2{{\textstyle{\frac{#1}{#2}}}}

\begin{document}
\title{Black hole remnants may exist if Starobinsky inflation occurred}

\author{Paul R. Anderson}\altaffiliation{\tt anderson@wfu.edu} \author{ Mathew J. Binkley} \author{Jillian M. Bjerke} \author{P. Wilson Cauley} \address{Department of Physics\\Wake Forest University\\Winston-Salem, NC  27109}

\begin{abstract}

Zero temperature black hole solutions to the semiclassical backreaction equations are investigated.  Evidence is provided that certain components of the stress-energy tensors for free quantum fields at the horizon only depend on the local geometry near the horizon.  This allows the semiclassical backreaction equations to be solved near the horizon.
It is found that macroscopic uncharged zero temperature black hole solutions to the equations may exist if the coefficient of one of the higher derivative terms in the gravitational Lagrangian is large enough and of the right sign for Starobinsky inflation to have occurred in the early Universe.

\end{abstract}
\maketitle

\section{Introduction}

With the discovery of black hole evaporation~\cite{hawking} came the fact that one can assign a temperature
to a black hole that is equal to the temperature of the thermal radiation that it emits.
This temperature is related to the surface gravity of the black hole.  Two questions which have still not been
resolved were raised by this discovery:  What is the end point of the evaporation process and what happens to the
information about how the black hole formed?
It may be that a fully quantum theory of gravity is necessary to answer these questions.  However, black hole solutions
to the four-dimensional semiclassical backreaction equations (SCE) have yet to be fully explored.  Thus it remains a possibility that
semiclassical gravity has something significant to say.

One possible answer to both the end point and information issues that has been suggested is that at late times the black hole evaporation process
may shut off leaving a zero temperature black hole remnant~\cite{chen}.  It is usually expected that such remnants would have sizes which are Planck scale and thus
would need to be described by a quantum theory of gravity.  It has been argued~\cite{giddings} that generically one might expect there to be an infinite amount of
pair production of such remnants if the information is stored inside those remnants.  However, it was also pointed out that there may be situations in which such infinite pair production does not occur.

Although it would be attractive to solve both the information and end point issues using black hole remnants, it is possible that they have separate solutions.  In that case
one can ask the question of whether such remnants could exist without being concerned about whether the information about how the black hole formed is inside of them.  Here we take this approach and investigate solutions to the semiclassical backreaction equations that correspond to static spherically symmetric zero temperature black holes (SZTBHs).  We focus on the solutions to the SCE near
the event horizons of such black holes and consider black holes with and without electric charges.  A macroscopic SZTBH must have an electric charge.  However, we find that if Starobinsky inflation~\cite{starobinsky,netto-et-al} occurred then the coefficient of one of the terms in the SCE is large enough and of the right sign so that it is possible to have uncharged SZTBH solutions to the SCE that are significantly larger than the Planck scale in size.

There is a long history of studying quantum effects in four-dimensional zero temperature black hole spacetimes.
The stress-energy for free massless quantized fields of spin $0$ and $\frac{1}{2}$ has been numerically computed in extreme Reissner-Nordstr\"{o}m (ERN) spacetimes
in four dimensions and found to be
regular on the event horizon~\cite{ahl,choag}.  It has also been analytically computed in Bertotti-Robinson spacetime~\cite{kaufman-sahni,ottewill-taylor} which becomes
a good approximation to the ERN geometry near the horizon.

SZTBH solutions to the linearized SCE in four dimensions were investigated for conformally invariant
fields in~\cite{aht,lowe} and for massive fields in~\cite{aht-mass,zara-mass}.  In~\cite{lowe,zara-mass} it was shown that solutions to the equations exist with different relationships between the mass of the black hole, the electric charge, and the radius of the event horizon
than occur for a classical ERN black hole.

The first solutions to the full nonlinear SCE in four dimensions that we are aware of which are relevant for SZTBHs, are for $AdS_2 \times S_2$ spacetimes in the case that a massless minimally coupled scalar field is present~\cite{solodukhin}.
This is the asymptotic form of the geometry near the event horizon of a SZTBH with a metric near the horizon that is of the same
general form as that for an ERN black hole near the horizon.  Both exact and approximate solutions were found, with the approximate ones being exact in certain limits.
It was found that solutions exist with no electric charge for a large range of values of the coefficient of the terms in the renormalized effective action for the scalar field that are quadratic in the curvature and local.

Constraints  on the behaviors of possible solutions to the full nonlinear SCE near the event horizons of SZTBHs were investigated in~\cite{am1,am2}.
       Assuming the usual higher derivative terms in the gravitational Lagrangian necessary for the renormalization of free quantum fields in curved space along with conformally invariant fields and a possible electric charge for the black hole, the trace of the SCE was solved near the horizon.  It was shown that there is a range of sizes for which no SZBHT solutions to the SCE are possible~\cite{am2}.  For metrics with power law behaviors for $g_{tt}$ and $g^{rr}$ near the horizon, constraints on the powers were obtained along with a relationship between the form of the metric near the horizon and the radius of the horizon~\cite{am1}.

  Here we continue the exploration of SZTBH solutions to the full nonlinear SCE.  We first add a constraint and then make
the argument that the most likely form of the metric near the horizon is one with $g_{tt}$ and $g^{rr}$ quadratic in $r-r_0$, with $r_0$ the radius of the event
horizon.  Next we consider metrics which have these forms near the horizon but different forms away from it. We compute the stress-energy tensors for massless
scalar fields with minimal and conformal coupling to the scalar curvature in these geometries.
Our results provide strong evidence that the values of the $\la {T_t}^t \ra$, $\la {T_r}^r \ra$, and $\la {T_\theta}^\theta \ra$ components on the horizon only depend on the geometry near the horizon.  This appears to be true for massless scalar fields with other couplings to the scalar curvature as well.  We expect that this property will also hold
for massless free fields of higher spin.  Our results provide evidence that the solutions in~\cite{solodukhin} can be used to describe the near horizon regions of SZTBH solutions to the SCE in the cases considered.

We have also computed the quantity
\be  \frac{\la {T_r}^r \ra - \la {T_t}^t \ra}{g_{tt}} \label{T-freely-falling} \ee
at the horizon.  This is related to the energy density seen by a freely falling observer who passes through the horizon.  If it diverges
at the horizon, then the observer sees an infinite energy density there.  We find that its value and in general the values of $\la {T_t}^t \ra_{,r}$ and
$\la {T_r}^r \ra_{,r}$ depend on the geometry away from the horizon as well as that near it.
We find that in some cases this quantity is finite on the horizon, but in many cases it is not.

We use our results to solve the SCE near the horizon when only conformally invariant fields are present along with the usual higher derivative
terms which are necessary for the renormalization of these fields.
Since the values of $\la {T_t}^t \ra$, $\la {T_r}^r \ra$, and $\la {T_\theta}^\theta \ra$ at the horizon depend only on the geometry near the horizon, we can
solve the SCE for the values of these components at the horizon.  If the stress-energy is finite on the horizon, then $\la {T_t}^t \ra = \la {T_r}^r \ra$ there and it suffices to solve the trace equation and the $rr$ component of the SCE.  Since the radial derivatives of these components depend upon the geometry away
from the horizon, we cannot say anything about SZTBH solutions to the SCE away from the horizon.  Therefore the
solutions we find tell us about the properties that physically acceptable SZTBH solutions to the SCE must have near the event horizon given the types of quantum fields
that we consider.

We restrict our attention to conformally invariant fields because most fields in the Standard Model
of particle physics are conformally invariant in the limit that their masses and interactions vanish. It was shown in~\cite{ahs} that the relevant quantity in determining the
importance of the mass is $m M$ in Planck units with $m$ the mass of the scalar field and $M$ the mass of the black hole.  For $m M \gtrsim 2$, the DeWitt-Schwinger approximation, which is a large mass approximation, was found to be valid.  Thus we expect the stress-energy tensor near the horizon to be approximately the same as that for a massless field if $ m M \ll 1$.

For the form of the metrics that we use, the results of~\cite{am1} for solutions to the trace equation indicate that there is a minimum size
that a SZTBH can have that is independent of the coefficients of the higher derivative terms in the equations.
Solving the $rr$ component of the SCE, we find that in many cases there is a more severe lower bound on the size that a SZTBH can have.
This lower bound corresponds to the case of zero electric charge and thus a solution satisfying this lower bound could serve as a black hole remnant.  If the coupling constant for the higher derivative term that leads to Starobinsky inflation~\cite{starobinsky,netto-et-al} has the right sign and magnitude for Starobinsky inflation to occur in
the early Universe~\cite{starobinsky-value}, and if it is significantly larger in magnitude than the other coupling constant, then the lower bound results in a black hole whose size is large enough compared with the Planck scale that semiclassical gravity can be valid.

In Sec.~\ref{sec:constraints} we review some results of~\cite{am1,am2} and come up with a new constraint on SZTBH solutions to the semiclassical backreaction equations.  In Sec.~\ref{sec:metrics} we argue that the most likely form for a zero temperature black hole metric near the horizon is given by~\eqref{fk-series-1}.  We also
show the specific form of the metric that we use for the numerical computations.  In Sec.~\ref{sec:num} we present some of our numerical results for components of the
stress-energy tensor in various candidate geometries.  Our solutions to the semiclassical backreaction equations near the horizon are given in Sec.~\ref{sec:solutions}.
Section~\ref{sec:summary} contains a summary and discussion of our results.
Throughout we use units such that $\hbar = c = G = k_B = 1$ and our sign conventions are those of~\cite{mtw}.

\section{Constraints on static spherically symmetric zero temperature black holes}
\label{sec:constraints}

In this section we first review constraints on the spacetime geometry near the event horizon of a SZTBH and then add a new constraint.

\subsection{Previous constraints}

Some constraints on the geometry of a SZTBH near the event horizon were obtained in~\cite{am1,am2} by
simply requiring that the components of the Riemann tensor in an orthonormal frame, or equivalently the Kretschmann scalar, be finite at the horizon.  Writing the metric in the form
\be ds^2 = -f(r)dt^2 + \frac{dr^2}{k(r)} + r^2 d\Omega^2  \;, \label{metric-1} \ee
one finds the surface gravity is
\be \kappa = \frac{v}{2} \sqrt{f k}  \;, \label{kappa} \ee
with
\be v \equiv \frac{f'}{f} \;. \label{v-def} \ee
Here primes denote derivatives with respect to $r$.
To have an event horizon it is necessary that $f = 0$ there and therefore that $v = \infty$.  To avoid a divergence of the Kretchmann scalar $R^{abcd} R_{abcd}$
at  the horizon it is necessary that $k = 0$ there as well.  To have a zero temperature black hole it is further necessary that $k' = 0$ at the horizon.
It is also necessary for zero temperature black holes that $k v^2$ be finite on the horizon and thus $k v = 0$ there.  Finally, for all zero
temperature black holes $\Box R$ cannot approach a constant on the horizon. It thus either diverges or vanishes there.

In~\cite{am1,am2} further constraints were obtained by considering conformally invariant quantum fields.  The trace of the stress-energy tensor for such fields is
the trace anomaly and is known in an arbitrary spacetime.  It is given in terms of the scalar curvature $R$, the Ricci tensor $R_{ab}$ and the Weyl tensor $C_{abcd}$ by
\be \langle T^q \rangle = \alpha \Box R + \beta \left(R_{a b}R^{a b} - \frac{1}{3} R^2 \right) + \gamma C_{abcd}C^{abcd} \;, \label{trace-anomaly} \ee
with
\bes
\bea \alpha &=& \frac{1}{2880 \pi^2} [N(0)+ 6 N(1/2) - 18 N(1)]  \;, \\
     \beta &=&   \frac{1}{2880 \pi^2} [N(0)+ 11 N(1/2) + 62 N(1)] \;, \\
      \gamma &=&   \frac{1}{2880 \pi^2} [N(0)+ \frac{7}{2} N(1/2) - 13 N(1)] \;. \eea \label{alp-bet-gam} \ees
Here $N(0)$, $N(1/2)$, and $N(1)$ are the numbers of conformally invariant scalar fields, four-component spin $1/2$ fields, and vector fields, respectively.
For the trace of the stress-energy tensor for a given type of conformally invariant field to be finite at the horizon of a SZTBH, it is clear that $\Box R$ cannot diverge there.  Thus since there is also the constraint mentioned above that $\Box R$ cannot be constant on the horizon, it is necessary that $\Box R = 0$ there.

Solutions to the semiclassical backreaction equations were investigated when only conformally invariant quantized fields are present.  The general form of these equations can be written as
\be G_{ab} = 8 \pi [T_{ab}^c + \langle T^q_{ab} \rangle +  h_1 \; ^{(1)}H_{ab} + h_2 \; ^{(C)}H_{ab}] \;, \label{SCE-1} \ee
where the superscripts $c$ and $q$ correspond to classical matter and quantum fields, respectively, and
\bes \bea  ^{(1)\!}H_{ab} &=& - \frac{1}{\sqrt{-g}} \frac{\delta}{\delta g^{ab}} \int d^4 x \, \sqrt{-g} \, R^2 = - 2 g_{ab} \Box R + 2 \nabla_a \nabla_b R - 2 R R_{ab}
+ \frc12 g_{ab} R^2 \;, \label{H1-def} \\
^{(C)\!}H_{ab} &=& - \frac{1}{\sqrt{-g}} \frac{\delta}{\delta g^{ab}} \int d^4 x \, \sqrt{-g} \, C_{abcd} C^{abcd} = -4 \nabla^c \nabla^d C_{acbd} - 2 R^{cd} C_{acbd} \;.
\label{HC-def} \eea \ees
The coefficients $h_1$ and $h_2$ are constants which must in principle be determined experimentally.

An important constraint was obtained from the trace of the SCE.  The only classical matter we consider here is the classical electric field that occurs if the black hole has an electric charge $Q$.  Since the electromagnetic field is conformally invariant, the trace of $T_{ab}^c$ is zero.  From~\eqref{HC-def} one sees that the trace of $^{(C)\!}H_{ab}$ is also zero due to its dependence on the Weyl tensor.  From~\eqref{H1-def} it is easily seen that $^{(1)\!}H_a^a = - 6 \Box R$.  Setting $\Box R = 0$ on the horizon gives
\be  -R = 8 \pi [ \langle T^q \rangle  ] \;. \label{SCE-trace} \ee
To derive the constraint the following  component of the Riemann tensor in an orthonormal frame was considered:
\be A(r) \equiv R_{\hat{t}\hat{r}\hat{t}\hat{r}} = \frac{v' k}{2} + \frac{v k'}{4} + \frac{v^2 k}{4} \;. \label{A-eq} \ee
Clearly this must be finite or there is a curvature singularity at the horizon.  By integrating one obtains
\be k = \frac{B_0}{v^2 f} + \frac{4}{v^2 f} \int_{r_0}^r f'(r_2) A(r_2) dr_2 \label{k1} \;, \ee
where $r_0$ is the radius of the event horizon.  Multiplying by $v^2 f$ and comparing with~\eqref{kappa}, one finds that, so long as $A_0 \equiv A(r_0)$ is finite on the horizon, $B_0 = 4 \kappa^2$.
Thus for the zero temperature black holes we are considering, $B_0 = 0$.  In~\cite{am2} these results were used to solve~\eqref{SCE-trace} on the horizon
with the result that
\bea A_0 &=& \frac{1}{16 \pi (\beta + 2 \gamma) r_0^2} \left[ 3 r_0^2 - 32 \pi (\beta-\gamma) \pm \left(768 \pi^2 \beta^2 - 3072 \pi^2 \beta \gamma - 288 \pi \beta r_0^2 + 9 r_0^4\right)^{1/2} \right]  \;. \eea
For physically acceptable solutions, $A_0$ must be real, which means there can be no solutions with $r_0$ in the range $r_{-} < r_0 < r_+$ with
\be r_\pm = 4 (\pi \beta)^{1/2} \left[1 \pm \left(\frac{2}{3 \beta} \right)^{1/2} (\beta + 2 \gamma)^{1/2} \right]^{1/2} \;. \label{size-gap}  \ee

\subsection{New constraints}

 A new constraint, which to our knowledge has not been presented elsewhere, can be obtained by first requiring that the curvature seen
by a freely falling observer in an orthonormal frame be finite.  In such a frame one
component of the Einstein tensor near the horizon depends in part upon the combination\footnote{This combination of components for the stress-energy tensor
is part of the energy density and pressure seen by a freely falling observer passing through the event horizon on a radial geodesic~\cite{lha-2D}.}
\begin{equation}
\frac{1}{f} (G_r^r - G_t^t) = \frac{k f'}{r f^2} - \frac{k'}{r f} \equiv - F(r) \; \label{eq:Grrtt}
\end{equation}
For the curvature to be finite at the horizon, it is clear that $F(r_0)$ must be finite.
This equation can be formally integrated with the result that
\be k = f \left[ a_1 + \int_{r_0}^r r_1 F(r_1) d r_1 \right] \;, \label{k2} \ee
where $a_1$ is an integration constant.
Equating ~\eqref{k1} and~\eqref{k2} and using the definition~\eqref{v-def} gives
\be (f')^2 = \frac{4 \int_{r_0}^r f'(r_2) A(r_2) dr_2}{a_1 + \int_{r_0}^r r_1 F(r_1) d r_1 }  \;. \ee
In~\cite{am2} it was shown that for SZTBH solutions to the SCE when only conformally invariant fields are present,
\be A_0  > 0  \;.  \label{A0}  \ee
Then we find that to leading order near the horizon
\be \frac{(f')^2}{f} = \frac{4 A_0}{a_1 + \int_{r_0}^r \, r_1 F(r_1) d r_1 } \;. \label{constraint1} \ee

Next we consider what this constraint implies for various values of $a_1$ and $F_0 \equiv F(r_0)$.
First it is necessary that $a_1 \ge 0$ since if $a_1 \ne 0$, then it dominates the denominator near the horizon.
If $a_1 >0$, then near the horizon
\be \frac{(f')^2}{f} = \frac{4 A_0}{a_1} \;. \ee
Integrating and using~\eqref{k1} gives
\bea f &=& \frac{A_0}{a_1} (r-r_0)^2  \;,  \nonumber \\
     k &=& A_0 (r-r_0)^2  \;. \label{f-k-quadratic} \eea

If $a_1 = 0$ and $F(r_0) = F_0 > 0$, then one can integrate~\eqref{constraint1} and use~\eqref{k1} to show that
\bes \bea f &=& \frac{4 A_0}{r_0 F_0} (r-r_0) \label{f-c10-F0} \\
     k &=& 4 A_0 (r-r_0)^2  \label{k-c10-F0} \eea \label{f-linear-k-quadratic} \ees

Finally if $a_1 = F_0 = 0$, then near the horizon
\be \frac{(f')^2}{f} = \frac{4 A_0}{\int_{r_0}^r \; r_2 F(r_2) d r_2 }   \;.  \label{constraint2} \ee
Taking the square root and integrating gives
\be 2 f^{1/2} = \int_{r_0}^r \, \left[ \frac{4 A_0}{\int_{r_0}^{r_3} \; r_2 F(r_2) d r_2 } \right]^{1/2} dr_3 \;. \ee
The minimum value of the right-hand side would occur if $F(r_0) >0$, and one would then obtain the result~\eqref{f-c10-F0}
for which $f'$ is constant at the horizon.  Thus $f'$ must have an infinite value on the horizon.  Further the function $F(r)$ cannot
vanish too rapidly as the horizon is approached or $f$ would not be equal to zero at the horizon.  As an example, suppose $f = a_4 (r-r_0)^p$ near the horizon
with $0 < p < 1$.  Then it is not hard to show that near the horizon
\bea k &=& \frac{4 A_0}{p^2} (r-r_0)^2 \;, \nonumber \\
     F(r) &=& \frac{4 A_0}{r_0 p^2 a_4} (r-r_0)^{1-p}  \;. \eea

\section{Metrics Considered Here}
\label{sec:metrics}

In the previous section constraints were found on the form of the metric for SZTBHs near the event horizon.
It was found that if only conformally invariant fields are present, then for SZTBH solutions to the SCE, metrics of the form~\eqref{f-k-quadratic}
and~\eqref{f-linear-k-quadratic} are allowed.  It was shown that for all other solutions $f' \rightarrow \infty$ at the horizon, which means there is no smooth way to continue $f$ across the horizon, and the coordinate system breaks down in a more significant way than it does for Schwarzschild or Reissner-Nordstr\"{o}m spacetimes.  If $f$ is linear and $k$ is quadratic at the horizon, then the obvious way of continuing $f$ and $k$ across the horizon leads to Euclidean space.  On the other hand if both $f$ and $k$ are quadratic near the horizon, then the metric is of the same form as that of an extreme Reissner-Nordstr\"{o}m spacetime.  From the point of view of the semiclassical backreaction equations, this is clearly the form of most interest and the one that will be pursued here.

In general the stress-energy tensor for a quantum field is a nonlocal quantity.  Therefore it is necessary to know the geometry everywhere in the causal past of a given spacetime point in order to compute the stress-energy tensor at that point.  For a SZTBH solution to the SCE outside the event horizon, this means knowing the geometry everywhere outside of the event horizon.  One can of course guess the geometry, but it is extremely unlikely that any guess would correspond to a solution to the SCE.  However, we have numerical evidence that for a SZTBH most components of the stress-energy tensor on the horizon depend only on the geometry near the horizon.  This allows us to solve the semiclassical backreaction equations near the horizon to determine that geometry.

Our conjecture concerns metrics for SZTBHs that near the event horizon have the leading order behaviors
\bes \bea f &\rightarrow& \left(\frac{r-r_0}{r_0}\right)^2 \;, \label{f-series-1} \\
         k &\rightarrow&  b_2 \left(\frac{r-r_0}{r_0}\right)^2 \;. \label{k-series-1}
\eea \label{fk-series-1} \ees
 Note that the coefficient for $f$ has been set to $1$ here because it is always possible to do this by rescaling the coordinate time $t$ in~\eqref{metric-1}.  The conjecture states that for a massless scalar field with arbitrary coupling to the scalar curvature, in SZTBH spacetimes for which $f$ and $k$  have the above form near the horizon, the values of the components $\la {T_t}^t \ra$, $ \la {T_r}^r \ra$, $\la {T_\theta}^\theta \ra$, and $ \la {T_\phi}^\phi \ra$ on the event horizon depend on the coefficient $b_2$, but not on the behaviors of $f$ and $k$ away from the horizon.

Previous work provides some evidence for this conjecture. In~\cite{ahl} it was shown numerically that on the event horizon of an extreme Reissner-Nordstr\"{o}m black hole
($b_2 = 1$) one finds that for a massless scalar field with arbitrary coupling to the scalar curvature
\be \la {T_t}^t \ra =  \la {T_r}^r \ra =  \la {T_\theta}^\theta \ra =  \la {T_\phi}^\phi \ra = \frac{1}{2880 \pi^2 M^4}  \;. \ee
It was also shown in~\cite{ahl} that these are the same values as those for the stress-energy tensor for the conformally coupled ($\xi = 1/6$) massless scalar field in the Bertotti-Robinson spacetime which is obtained by expanding the extreme Reissner-Nordstr\"{o}m metric in a series about $r = r_0$
and keeping only the lowest order terms.  In Sec.~\ref{sec:num} we give a more technical explanation of why the conjecture works along with numerical results for another value of $b_2$ that support it.

If the conjecture is correct then the following procedure will work to solve the semiclassical backreaction equations near the horizon.
Choose metric functions which approach~\eqref{fk-series-1} near the horizon for various values of $b_2$ and which have any convenient form away from it.
Then compute the stress-energy tensors for the quantum fields and evaluate their components at the horizon.  Next evaluate the left-hand sides of the trace and $rr$ components of the SCE.  They  depend only on $r_0$ and $b_2$ at the horizon.  Finally, since the ERN black hole has an electric charge,
include on the right-hand side of the SCE the classical electromagnetic stress-energy tensor that occurs if the black hole has an electric charge $Q$.
Then the trace of the SCE will be independent of $Q$ and should yield a relationship between $r_0$ and $b_2$.  The $rr$ component should yield a relationship between $r_0$, $b_2$, and $Q^2$.  Thus, for any desired size for the black hole, one could find the magnitude of the resulting
electric charge and the leading order behavior of the metric near the horizon.

In this paper we use the dimensionless radial coordinate
\be  s \equiv \frac{r - r_0}{r_0}  \;, \label{s-def} \ee
 and consider metrics of the following form near the event horizon:
\bes \bea f &=& a_2 s^2 + a_3 s^3 + \ldots \;, \label{f-series-2} \\
         k &=&  b_2 s^2 + b_3 s^3 + \ldots \;. \label{k-series-2}
\eea \label{fk-series-2} \ees
Note that without loss of generality we can absorb the value of the coefficient $a_2$ into the definition of the time coordinate $t$.  We do this for the computations
discussed here.

To compute the components of the stress-energy tensor it is necessary to specify the metrics everywhere outside of the event horizon.  So the actual metrics we consider are of the general form
\bes \bea f &=&\frac{ s^2}{(s+1)^2}  + \frac{s^3}{(s+1)^3} \; A_{33}  + \ldots \;,  \label{f-series-3} \\
  k &=&  \frac{s^2}{(s+1)^2} \; b_2  +  \frac{s^3}{(s+1)^3}\; B_{33}+ \ldots \label{k-series-3}  \eea \label{fk-series-3} \ees
Note that for an asymptotically flat spacetime $f \to c$ for some constant $c >0$ and $ k \to 1$ as $s \to \infty$.  The first condition is automatically satisfied by these
metrics.  For the second
\be  b_2 + B_{33} + \ldots = 1 \;. \ee
Since $b_2 >1$ it is necessary that at least one of the other terms in the sum be negative.

It is tempting to
make the conjecture that the first radial derivatives of the components of the stress tensor at the horizon
depend only on the values of $r_0, b_2, a_3$, and $b_3$.  However we have found that this is not the case.
 Thus it appears that this approach only allows one to find the behaviors of solutions to the SCE when its trace, $rr$, and $tt$ components
 are evaluated at the horizon.

\section{Numerical Results}
\label{sec:num}

We begin with a constraint on two components of the stress-energy tensor at the horizon.
The radial component of the conservation equation $\la {T_a}^b\ra_{;b} = 0$ is
\be {\la {T_r}^r}\ra_{,r} + \frac{1}{2 f} \frac{d f}{dr} (\la {T_r}^r\ra -\la {T_t}^t\ra) + \frac{2}{r}( \la {T_r}^r\ra - \la {T_\theta}^\theta \ra) = 0  \;. \label{Tcons} \ee
Note that since we consider only states which respect spherical symmetry, $\la {T_\phi}^\phi\ra = \la {T_\theta}^\theta \ra$.
For the metrics we consider $f^{-1} \frac{df}{dr} \sim (r-r_0)^{-1}$ near the horizon.  Thus for $\la {T_r}^r \ra_{,r}$ to be finite at the horizon it is
necessary that $\la {T_t}^t \ra = \la {T_r}^r \ra$ there.  This result is well known and our numerical results confirm that for the vacuum state this condition is always satisfied.

In the previous section, a conjecture was presented which states that for a massless scalar field the components $\la {T_t}^t \ra$, $ \la {T_r}^r \ra$, $\la {T_\theta}^\theta \ra$, and $ \la {T_\phi}^\phi \ra$ depend only on $r_0$ and the metric parameter $b_2$ when the metric is of the form~\eqref{fk-series-1} near the horizon.
It is possible to show using the general static spherically symmetric form of the expressions for $\la {T_a}^b \ra$~\cite{ahs}, the definition
\be  r = r_0 (1+ s) \;, \label{s-def} \ee
and the scaling $\omega \to \omega/r_0$, that the entire $r_0$ dependence for each of these components is $r_0^{-4}$.

In this section we first discuss the computation of these components on the horizon.  In the process we provide a technical explanation for why the conjecture should be correct.  Then we present the results of some of our numerical computations.

The method we use to compute the stress-energy tensor for a massless scalar field in a SZTBH spacetime is given in detail in~\cite{ahs}.  In this approach the mode equation in the Euclidean space associated with the exterior region of the black hole is solved.  For each value of the frequency $\omega$ and the angular momentum parameter $\ell$ there are two linearly independent solutions.  One of them we call $p_{\omega \ell}$, and it is regular at the horizon\footnote{There can be spacetimes where there are exceptions to this for small values of $\omega$.  However, in these cases the divergence is still less strong than for $q_{\omega \ell}$ at the horizon.} but diverges at infinity.  The other we call $q_{\omega \ell}$.  It is well behaved at infinity but diverges at the horizon.  The two-point function $\la \{\phi(x),\phi(x')\} \ra$ is a sum and integral over the product of these two mode functions.  The unrenormalized stress-energy tensor involves spacetime derivatives of the two-point function.

The fact that the stress-energy tensor depends only on the geometry near the horizon for
ERN spacetimes and our conjecture that this is the case in general for SZTBHs can be understood in two different ways.  First, it is easy to show that the proper distance to the horizon
along a radial spacelike geodesic from any point outside of it is infinite~\cite{am2}.  In Euclidean
space the distance is infinite for any path from outside the horizon to the horizon.  Thus
it makes sense qualitatively that the stress-energy tensor might depend only on the geometry
near the horizon.

 From a more technical point of view it is found that to leading order near the horizon $p_{\omega \ell}$ and $q_{\omega \ell}$ have exponential
factors of the form
\begin{equation}
 \exp(\pm \omega/(r-r_0)) \;.
\end{equation}
Since the boundary conditions for $q_{\omega \ell}$ are
fixed away from the horizon, changing these conditions simply amounts to adding some
part of the $p_{\omega \ell}$ mode to the original $q_{\omega \ell}$ mode.  Then a product of the $p_{\omega \ell}$ and $q_{\omega \ell}$ modes
simply results in the original product plus a term which is damped exponentially as the
horizon is approached.  Therefore it is plausible that in the limit that the horizon
is approached this exponentially damped term does not contribute to leading order to the mode sum that
makes up the stress-energy tensor.

The method in~\cite{ahs} allows us to compute the components of the stress-energy tensor anywhere outside the event
horizon.  The results can be extrapolated to the horizon.
There is a well-known ambiguity which occurs for the value of $\la T_{ab} \ra$ which comes from the renormalization counterterms.  For the conformally invariant scalar field this
results in a finite renormalization of the parameter $h_2$ in the semiclassical backreaction equations~\eqref{SCE-1}.  For the massless minimally coupled field it results
in finite renormalizations of both $h_1$ and $h_2$.  For the method we use there is an arbitrary constant in one term of the stress-energy tensor which is multiplied by $^{(C)}H_{ab}$ in the case of the conformally invariant field and which is multiplied by a linear combination of $^{(1)}H_{ab}$ and $^{(C)}H_{ab}$ for the massless minimally coupled scalar field. More details are given in~\cite{ahs}.  For the numerical results shown we chose the value of this constant to be zero.

\begin{figure}
\includegraphics[width=5in]{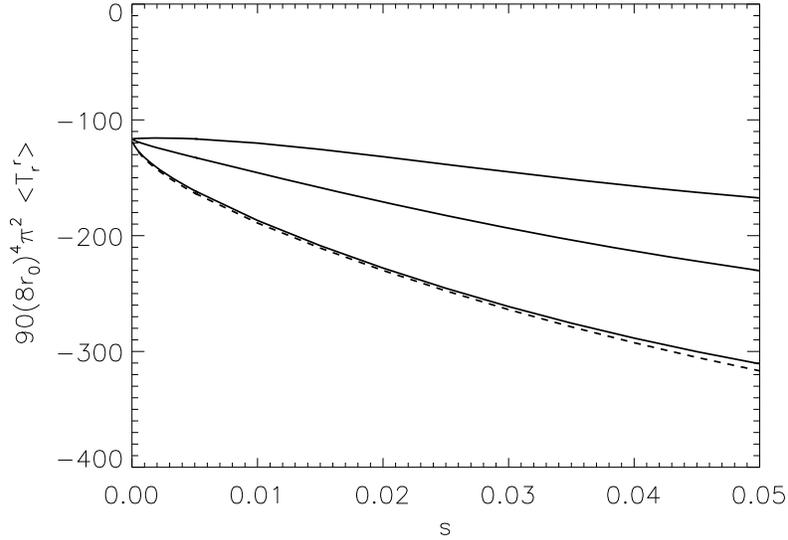}
\caption{\label{Fig:Trr} The quantity $\la {T_r}^r \ra$ is plotted near the horizon for a massless scalar field with $\xi = \frac{1}{6}$ when $b_2 = 2$.  All of the curves have the series~\eqref{fk-series-3} truncated at $A_{33}$ and $B_{33}$.  The solid curves have  $A_{33} = 0$ and thus $a_3 = -2$.  From top to bottom they have $B_{33} = 0$ ($b_3 = -4$), and $B_{33} = 1$  ($b_3 = -3$),
$B_{33} = 2$  ($b_3 = -2$).  The dashed curve has $A_{33} = B_{33} = 2$  ($a_{33} = 0$, $b_3 = -2$).
 }
\end{figure}

The field is conformally invariant if it is massless and $\xi = \frac{1}{6}$.  In this case the $\la {T_\theta}^\theta \ra$ component on the horizon is
related through the trace anomaly with the $\la {T_r}^r \ra$ component; see the next section for details.
Some of our results for $\la {T_r}^r \ra$ are shown in Fig.~\ref{Fig:Trr}.

\subsection{Results for another component}

\begin{figure}
\includegraphics[width=5in]{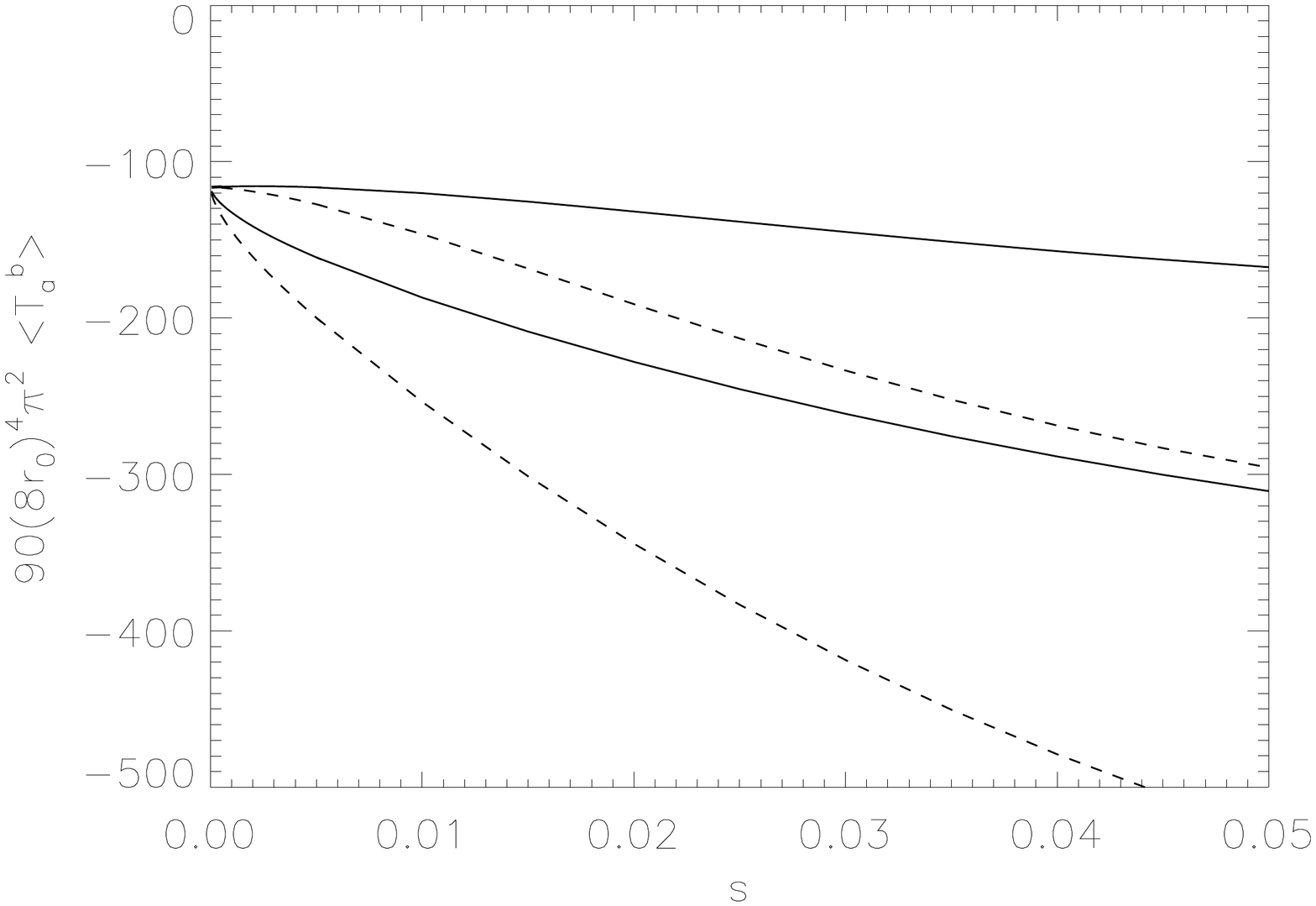}
\caption{\label{Fig:TrrTtt16} Components of the stress-energy tensor are plotted near the horizon for  a massless scalar field with $\xi = \frac{1}{6}$ when $b_2 = 2$.  All of the curves have the series~\eqref{fk-series-3} truncated at $A_{33}$ and $B_{33}$.  The upper solid and dashed curves show $\la {T_r}^r \ra$ and $\la {T_t}^t \ra$, respectively, for $A_{33} = B_{33} = 0$ ($a_3 = -2$, $b_3 = -4$).  The lower solid curve and lower dashed curve show $\la {T_r}^r \ra$ and $\la {T_t}^t \ra$,  respectively, for $A_{33} = 0$ and $B_{33} = 2$   ($a_3 = b_3 = -2$).
Note that in both cases the slope of the curve for $\la {T_r}^r \ra$ near the horizon is different from that of the curve for $\la {T_t}^t \ra$.
 }
\end{figure}

\begin{figure}
\includegraphics[width=5in]{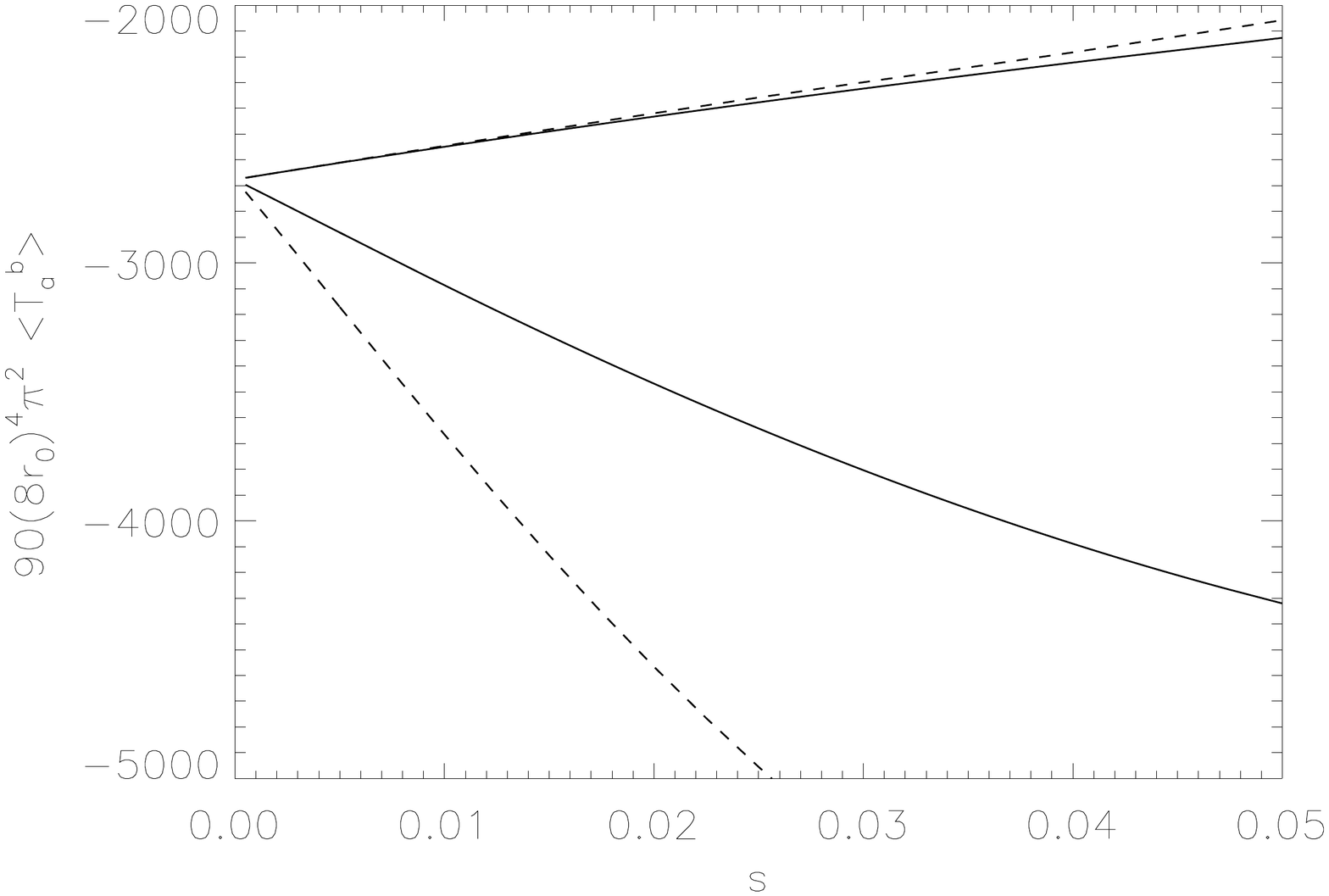}
\caption{\label{Fig:TrrTtt0} Components of the stress-energy tensor are plotted near the horizon for a massless scalar field with $\xi = 0$ when $b_2 = 2$.  All of the curves have the series~\eqref{fk-series-3} truncated at $A_{33}$ and $B_{33}$.  The upper solid and dashed curves show $\la {T_r}^r \ra$ and $\la {T_t}^t \ra$, respectively, for $A_{33} = B_{33} = 0$ ($a_3 = -2$, $b_3 = -4$).  The lower solid curve and lower dashed curve show $\la {T_r}^r \ra$ and $\la {T_t}^t \ra$,  respectively, for $A_{33} = 4$ and $B_{33} = 6$   ($a_3 = b_3 = 2$).
Note that the slopes of the upper curves approach each other near the horizon but that for the lower curves
the slope of the curve for $\la {T_r}^r \ra$ near the horizon is different from that of the curve for $\la {T_t}^t \ra$.
 }
\end{figure}

Computation of the stress-energy tensor in an orthonormal frame attached to a freely falling observer moving in the radial direction shows that
as the observer falls through the horizon, the observer will observe an infinite stress-energy unless $\la {T_t}^t \ra$, $\la {T_r}^r \ra$, and $g^{-1}_{tt} (\la {T_r}^r\ra - \la {T_t}^t\ra)$
are all finite on the horizon~\cite{lha-2D}.   Since $g_{tt} \sim (r-r_0)^2$, this component is divergent unless $\la {T_t}^t\ra_{,r} = \la {T_r}^r\ra_{,r}$ on the horizon.
From Fig.~\ref{Fig:TrrTtt16} it is clear that this is not the case for all geometries of the form~\eqref{fk-series-3}.  In fact we have not found an example where this condition
is satisfied for conformal coupling $\xi = \frac{1}{6}$.  However, as shown in Fig.~\ref{Fig:TrrTtt0} we have found examples where it appears to be satisfied for minimal coupling $\xi = 0$.

The values of $\la {T_t}^t \ra$ and $\la {T_r}^r \ra$ on the horizon depend only on the geometry near the horizon and in particular on the values of $b_2$ and $r_0$.  However, some of our numerical results indicate that the values of
 $\la {T_t}^t \ra_{,r}$ and $\la {T_r}^r \ra_{,r}$ at the horizon appear to depend on the geometry away from the horizon as well.  Thus it is quite possible that there
 are spacetime geometries for which $\la {T_t}^t\ra_{,r} = \la {T_r}^r\ra_{,r}$ on the horizon for $\xi = \frac{1}{6}$.

\section{Solutions to the semiclassical backreaction equations near the horizon}
\label{sec:solutions}

In this section we solve the semiclassical backreaction equations near the horizon using our results in Sec.~\ref{sec:num} which assume a metric of the form~\eqref{fk-series-3}.  We begin by reviewing the solution to the trace equation.
If only conformally invariant quantum fields are present, then the trace equation is given by substituting~\eqref{trace-anomaly} into the trace of~\eqref{SCE-1}.  Evaluating at the horizon and recalling that $\Box R = 0$ there, one finds that the resulting equation can be solved for $r_0$ as a function of $b_2$ with the result that
\be r_0^2 = \frac{\pi}{3 (b_2-1)} \left[ 8 (\beta+2 \gamma)(b_2^2+1) + 32 (\beta-\gamma) b_2 \right]
 \;. \label{SCE-trace-hor} \ee
It is easy to show from~\eqref{alp-bet-gam} that $\beta+ 2\gamma > 0$ and $\beta - \gamma \ge 0$.  Thus physically acceptable solutions only exist if $b_2 > 1$.  It is worth noting that for an ERN black hole, $b_2 = 1$.  Thus the ERN solution to the classical Einstein equations is not a solution
to the SCE if only conformally invariant fields are present.

There is a minimum radius which occurs for
\be (b_2)_{\rm min} = 1 + \sqrt{\frac{6 \beta}{\beta + 2 \gamma}}  \;. \label{b2-r0min} \ee
It is
\be (r_0^2)_{\rm min} = 16 \pi \left( \sqrt{\frac{2}{3}} \sqrt{\beta (\beta+2 \gamma)} + \beta \right)  \;. \label{r0-min} \ee
For the Standard Model $N_0 = 4$, $N_{1/2} = 45$, and $N_1 = 12$, so
\bes \bea \beta &=& \frac{1243}{2880 \pi^2} \;, \label{bet-std-model} \\
     \gamma &=& \frac{11}{5760 \pi^2} \;,\label{gam-std-model} \eea \label{bet-gam-std-model} \ees
and
\bea (b_2)_{\rm min} & =& 1+ \sqrt{\frac{113}{19}} \approx 3.4  \, \nonumber \\
      (r_0)_{\rm min} & = & \frac{1}{\sqrt{\pi}} \sqrt{\frac{1243}{180}+ \frac{11\sqrt{2147}}{90}}  \approx 2.0 \;. \label{b2-r0-min-std-model} \eea
Thus for the Standard Model the minimum size is of order the Planck length.  However there are many more particles in Grand Unified Theories, so
 the minimum size could be significantly larger than the Planck scale.
Further this minimum is really only a constraint because it comes from just one of the backreaction equations.  The actual minimum could be larger.  Note that it does not depend
on the coefficients $h_1$ and $h_2$ of the higher derivative terms in the semiclassical backreaction equations nor does it depend on the charge of
the black hole.

Continuing the analysis of the solutions to the trace equation at the horizon, for small values of $b_2 - 1 >0$, the radius is
\be r_0^2 \approx \frac{16 \pi \beta}{b_2-1} \;. \label{r0-b2-1} \ee
and the scalar curvature at the horizon is
\be R = -\frac{2 (b_2-1)}{r_0^2}  \approx -\frac{(b_2-1)^2}{8 \pi \beta}    \;. \label{R-b2-1}  \ee
Thus for $1 < b_2 \le (b_2)_{\rm min}$ the size of the event horizon ranges from infinity to its minimum value and the scalar curvature is small when the horizon
size is large.  Thus these values of $b_2$ result in physically acceptable solutions.

For very large values of $b_2$,
\be r_0^2 \approx \frac{8 \pi}{3} (\beta + 2 \gamma) b_2   \;, \ee
and
\be R \approx -\frac{3}{4 \pi (\beta + 2 \gamma)}  \;. \ee
Since $R$ does not get small as $r_0$ gets large, the solutions with $b_2 \gg (b_2)_{\rm min}$  are probably not physically acceptable.

To go further we examine the ``rr'' component of the semiclassical backreaction equations.  At the horizon the equation is
\be -\frac{1}{r_0^2} = 8 \pi \left[-\frac{Q^2}{8 \pi r_0^4} +  ({T_r}^r)_0 - \frac{2}{r_0^4} (b_2^2-1) \left(\frac{h_2}{3} + h_1\right)\right] \;. \ee
Here $({T_r}^r)_0$ is the value of $\la {T_r}^r \ra$ evaluated at $r = r_0$.
Thus the charge of the black hole which corresponds to a given value of $r_0$ and hence $b_2$ is
\be Q^2 = r_0^2 + 8 \pi \left[ r_0^4 ({T_r}^r)_0 - 2 (b_2^2-1) \left(\frac{h_2}{3} + h_1\right)\right]  \;. \label{Q2} \ee
Note that $r_0^4 ({T_r}^r)_0$ depends on $b_2$ and not $r_0$.  Thus this equation gives a relationship between the charge $Q$, the radius
$r_0$, and the metric parameter $b_2$ for fixed values of $h_1$ and $h_2$.

It is of interest to see whether it is possible to have $Q = 0$.  Since $b_2 >1$, it is clearly not possible if $h_2 + 3 h_1 < 0$ and $({T_r}^r)_0 > 0$.
Even for values of these quantities where it is possible to have $Q = 0$, the resulting radius of the black hole will be of the Planck scale or smaller
unless there is a large number of fields and $({T_r}^r)_0 < 0$ and/or $h_1 + h_2/3 \gg 1$.  The latter condition can be satisfied if $h_2 \gg - h_1$ and the Universe underwent
Starobinsky inflation, which requires $h_1 \sim 10^9$~\cite{starobinsky-value}.

If $Q^2 = 0$, then~\eqref{Q2} gives a second equation for $r_0$.  Combining~\eqref{SCE-trace-hor} and~\eqref{Q2} gives
\be  (\beta+2 \gamma)(b_2^2+1) + 4 (\beta-\gamma) b_2 + 3 r_0^4 ({T_r}^r)_0 (b_2-1) - 6 \left(\frac{h_2}{3}+h_1 \right) (b_2+1) (b_2-1)^2  = 0 \;. \label{b2-eq} \ee
For a black hole much larger than the Planck scale, Eq.~\eqref{SCE-trace-hor} implies that $b_2 \approx 1$, which in turn implies that
 the metric near the horizon is nearly the same as that of the extreme Reissner-Nordstr\"{o}m metric.
In that case one expects $({T_r}^r)_0$ to be approximately equal
to its value in an ERN spacetime which for conformally invariant fields is\footnote{The ERN value for massless scalar fields was computed in~\cite{ahl} and
for the spin $\frac{1}{2}$ was computed in~\cite{choag}.  In~\cite{aht} it was argued from the conformal invariance of the spacetime near the horizon that
for conformally invariant fields in general it is given by~\eqref{Trr0}.}
\be ({T_r}^r)_0 = \frac{\beta}{r_0^4} \;. \label{Trr0} \ee
Using this as an approximation for $({T_r}^r)_0$ in~\eqref{b2-eq} along with $b_2 \approx 1$, and assuming $|h_2| \ll h_1$ gives
\be (b_2-1)^2 = \frac{\beta}{2 h_1}  \;. \label{b2m1} \ee
Substituting this into~\eqref{SCE-trace-hor} gives
\be r_0 \approx    ( 512 \pi^2 \beta h_1)^{1/4}  \;. \label{r0-remnant} \ee
Using $10^9$ for $h_1$ and the value of $\beta$ for the Standard Model~\eqref{bet-gam-std-model} gives $r_0 \approx 700$, which is well above the Planck scale where $r_0 \sim 1$.
For grand unified theories $\beta$ and hence $r_0$ are even larger.  Thus if Starobinsky inflation occurred it is possible that black hole remnants could exist that are compatible with and predicted by semiclassical gravity.

\section{Summary and Conclusions}
\label{sec:summary}

We have examined constraints on the form of the metric for SZTBH solutions to the semiclassical backreaction equations and found that the most likely form the metric would take is that both $g_{tt}$ and $g^{rr}$ are quadratic in $r-r_0$ near the horizon.  Restricting our attention to metrics of this
form, we have numerically computed the stress-energy tensor for both the conformally invariant scalar field and the massless minimally coupled scalar field in spacetimes with
metrics of the form~\eqref{fk-series-3}.  It has been found in all cases considered that the value of $\la {T_t}^t \ra = \la {T_r}^r\ra$ on the horizon depends only on the metric parameter $b_2$ and on the radius $r_0$ of the event horizon.  This makes it possible to determine the leading order behaviors of solutions to the SCE near the horizon.

We have examined the solutions to the SCE near the horizon when only conformally invariant quantum fields are present.  It was shown in~\cite{ahs} that for a massive scalar
field the large mass condition is given by $m M \sim 2$ in Planck units.  For the small mass limit ($m M \ll 1$), most massive free fields are approximately conformally invariant. For small enough black holes this includes most of the fields in the Standard Model if their interactions can be neglected.
We have found that near the horizon zero temperature solutions to the SCE can exist even if the black hole has no electric charge.
Of course only knowing their behaviors near the horizon does not guarantee that these solutions have physically realistic geometries
far from the horizon and that they could thus correspond to realistic zero temperature black holes.  Even if the
geometries are physically realistic,  it does not guarantee that the black hole evaporation process really does shut off at late times when
the black hole is small and therefore that black hole remnants exist. What one does expect however, is that backreaction effects due to quantum
fields should be larger for black holes of smaller sizes.  Thus it is possible that such effects could result in progressively smaller surface gravities and
hence progressively lower temperatures for such black holes with the limit being the uncharged SZTBH solutions discussed here.

The $r r$ component of the semiclassical backreaction equations provides a relation between $b_2$, $r_0$, and the black hole charge $Q$ along with the coefficients $h_1$ and $h_2$ of the $R^2$ and Weyl squared terms in the gravitational Lagrangian.
If only conformally invariant fields are present, we have shown that this relationship allows for an electric charge of zero for the black hole
 if $\la {T_r}^r \ra$ on the horizon has a large enough negative value and/or $h_2/3 + h_1$ has a large enough positive value.
For values of $|h_1|$ and $|h_2|$ less than or of order unity there
would need to be an enormous number of quantum fields for the corresponding black hole to be larger in size than the Planck scale.  However, if Starobinsky inflation occurred so that $h_1 \sim 10^9$, and if $h_2 \gg - h_1$, then zero temperature black holes with sizes significantly above the Planck scale could exist even for the number of quantum fields in the Standard Model.  Thus if Starobinsky inflation occurred, then it is possible that black hole remnants could exist that are large enough that semiclassical gravity could be used to describe them.  As such they could provide an answer to the question of what the end state of the black hole evaporation process is.

\acknowledgments

P.R.A. would like to acknowledge helpful discussions with William Hiscock, Jacob Bekenstein, and Eric Carlson.  He would like to thank Los Alamos National Laboratory, the Racah Institute of Physics, and the University of Valencia for hospitality.  P.R.A. also acknowledges the Einstein Center at Hebrew University of Jerusalem,
the Forchheimer Foundation, and the Spanish Ministerio de Educaci\'{o}n y Ciencia for financial support.
This work was supported in part by the National Science Foundation under Grants No. PHY-9800971, No. PHY-0070981, No. PHY-0556292, No. PHY-0856050, No. PHY-1308325, and No. PHY-1505875.

\end{document}